\documentclass[12pt]{iopart}
\expandafter\let\csname equation*\endcsname\relax
\expandafter\let\csname endequation*\endcsname\relax
\usepackage{graphicx,bm,citesort,amsmath,iopams,color}

\newcommand{\AnM}{ {\it Annals Math. }}
\newcommand{\APB}{ {\it Ann. Phys. (Berlin) }}
\newcommand{\CMP}{ {\it Commun. Math. Phys. }}

\newcommand{\CPl}{ {\it Chem. Phys. Lett. }}
\newcommand{\CRA}{ {\it C. R. Acad. Sci. Ser.} A }

\newcommand{\EPJD}{ {\it Eur. Phys. J.} D }
\newcommand{\EPJP}{ {\it Eur. Phys. J. Plus }}

\newcommand{\IJQC}{ {\it Int. J. Quantum Chem. }}

\newcommand{\JPCo}{ {\it J. Phys. Commun. }}

\newcommand{\PA}{ { \it Physica} A }
\newcommand{\PLA}{ \PL A }

\newcommand{\PRA}{ \PR A }
\newcommand{\PRB}{ \PR B }

\newcommand{\SLM}{ {\it Superlattice Microst. }}

\begin{document}
\title[Quantum-information theory of magnetic field influence on circular dots]{Quantum-information theory of magnetic field influence on circular dots with different boundary conditions}
\author{H Shafeekali$^2$ and O Olendski$^{1,2}$}
\address{$^1$ Atomic and Molecular Engineering Laboratory, Belarusian State University, Skarina Ave. 4, Minsk 220050, Belarus}
\address{$^2$ Department of Applied Physics and Astronomy, University of Sharjah, P.O. Box 27272, Sharjah, United Arab Emirates}
\ead{oolendski@sharjah.ac.ae}

\begin{abstract}
Influence of the transverse uniform magnetic field $\bf B$ on position (subscript $\rho$) and momentum ($\gamma$) Shannon quantum-information entropies $S_{\rho,\gamma}$, Fisher informations $I_{\rho,\gamma}$ and informational energies  $O_{\rho,\gamma}$ is studied theoretically for the 2D circular quantum dots (QDs) whose circumference supports homogeneous either Dirichlet or Neumann boundary condition (BC). Comparative analysis reveals similarities and differences of the influence on the properties of the structure of the surface interaction with the magnetic field. A conspicuous distinction between the two spectra are crossings at the increasing induction of the Neumann energies with the same radial quantum number $n$ and adjacent non-positive angular indices $m$. At the growing $B$, either system undergoes Landau condensation when its characteristics turn into their uniform field counterparts. It is shown that for the Dirichlet system this transformation takes place at the smaller magnetic intensities; e.g., the Dirichlet sum $S_{\rho_{00}}+S_{\gamma_{00}}$ on its approach from above to a fundamental limit $2(1+\ln\pi)$ is at any $B$ smaller than the corresponding Neumann quantity what physically means that the former geometry provides more total information about the position and motion of the particle. It is pointed out that the widely accepted disequilibrium uncertainty relation $O_\rho O_\gamma\leq(2\pi)^{-\mathtt{d}}$, with $\mathtt{d}$ being a dimensionality of the system, is violated by the Neumann QD in the magnetic field. Comparison with electrostatic harmonic confinement is performed; in particular, contrary to the hard-wall QDs, for this configuration the sums $S_\rho+S_\gamma$ and the products $I_\rho I_\gamma$ and  $O_\rho O_\gamma$ do not depend on the field. Physical interpretation is based on the different roles of the two BCs and their interplay with the field: Dirichlet (Neumann) surface is a repulsive (attractive) interface.
\end{abstract}
\vskip.7in

\noindent

\submitto{\PS}
\maketitle

\section{Introduction}\label{Sec_Intro}
Despite of a long history of research, quantum dots (QDs) continue to attract a careful attention of physicists, mathematicians, chemists and scientists from other closely related fields \cite{Cotta1}; e.g., recent numerical and theoretical analysis \cite{Olendski1} compared an influence of the boundary conditions (BCs) on quantum-information measures, such as Shannon entropy \cite{Shannon1,Shannon2}, Fisher information \cite{Fisher1,Frieden1} and informational energy \cite{Onicescu1}, of the two-dimensional (2D) circular QDs. Mentioned above functionals of the position $\bf r$
\begin{subequations}\label{Densities1}
\begin{align}\label{Densities1_R}
\rho_\mathtt{n}({\bf r})&=|\Psi_\mathtt{n}({\bf r})|^2
\intertext{and momentum $\bf p$ (or, more correctly, wave vector ${\bf k}\equiv{\bf p}/\hbar$)}
\gamma_\mathtt{n}({\bf k})&=|\Phi_\mathtt{n}({\bf k})|^2
\end{align}
\end{subequations}
densities quantitatively describe different aspects of the particle distribution in the corresponding $\mathtt{d}$D space, with associated functions $\Psi_\mathtt{n}({\bf r})$ and $\Phi_\mathtt{n}({\bf k})$ being Fourier transforms of each other:
\begin{subequations}\label{Fourier1}
\begin{align}\label{Fourier1_1}
\Phi_\mathtt{n}({\bf k})&=\frac{1}{(2\pi)^{\mathtt{d}/2}}\int_{\mathcal{D}_\rho^{(\mathtt{d})}}\Psi_\mathtt{n}({\bf r})e^{-i{\bf kr}}d{\bf r},\\
\label{Fourier1_2}
\Psi_\mathtt{n}({\bf r})&=\frac{1}{(2\pi)^{\mathtt{d}/2}}\int_{\mathcal{D}_\gamma^{(\mathtt{d})}}\Phi_\mathtt{n}({\bf k})e^{i{\bf rk}}d{\bf k},
\end{align}
\end{subequations}
$\mathtt{n}=1,2,3,\ldots$, and $\mathcal{D}_\rho^{(\mathtt{d})}$ and $\mathcal{D}_\gamma^{(\mathtt{d})}$  are the domains where $\Psi_\mathtt{n}({\bf r})$ and $\Phi_\mathtt{n}({\bf k})$, respectively, are defined. In addition, each set of waveforms is orthonormalized:
\begin{equation}\label{OrthoNormality1}
\int_{\mathcal{D}_\rho^{(\mathtt{d})}}\Psi_\mathtt{n'}^\ast({\bf r})\Psi_\mathtt{n}d{\bf r}=\int_{\mathcal{D}_\gamma^{(\mathtt{d})}}\Phi_\mathtt{n'}^*({\bf k})\Phi_\mathtt{n}({\bf k})d{\bf k}=\delta_\mathtt{n'n},
\end{equation}
with $\delta_\mathtt{n'n}$ being a Kronecker delta, $\mathtt{n'}=1,2,3,\ldots$. Dependencies $\Psi_\mathtt{n}({\bf r})$ are eigen functions of the one-particle $\mathtt{d}$D Schr\"{o}dinger equation
\begin{equation}\label{Schrodinger1}
\widehat{H}^{(\mathtt{d})}\Psi_\mathtt{n}({\bf r})=E_\mathtt{n}\Psi_\mathtt{n}({\bf r})
\end{equation}
with the Hamiltonian
\begin{equation}\label{Hamiltonian1}
\widehat{H}^{(\mathtt{d})}=\frac{1}{2M}\left[i\hbar{\bm\nabla}_{\bf r}+q{\bf A}({\bf r})\right]^2+V({\bf r}),
\end{equation}
where $V({\bf r})$ and ${\bf A}({\bf r})$ are the external electrostatic and vector potentials, respectively, in which the particle with  mass $M$, charge $q$ and quantized energy $E_\mathtt{n}$ is moving. Magnetic field ${\bf B}({\bf r})$ enters the above equation via the vector potential, ${\bf B}({\bf r})={\bm\nabla}_{\bf r}\times{\bf A}({\bf r})$.

Mathematical expressions for the position $S_{\rho_\mathtt{n}}$ and momentum $S_{\gamma_\mathtt{n}}$ Shannon quantum-information entropies
\begin{subequations}\label{Shannon1}
\begin{align}\label{Shannon1_R}
S_{\rho_\mathtt{n}}&=-\int_{\mathcal{D}_\rho^{(\mathtt{d})}}\rho_\mathtt{n}({\bf r})\ln\rho_\mathtt{n}({\bf r})d{\bf r}\\
\label{Shannon1_K}
S_{\gamma_\mathtt{n}}&=-\int_{\mathcal{D}_\gamma^{(\mathtt{d})}}\gamma_\mathtt{n}({\bf k})\ln\gamma_\mathtt{n}({\bf k})d{\bf k}
\end{align}
\end{subequations} 
physically express the amount of information about the particle position, equation~\eref{Shannon1_R}, or motion, equation~\eref{Shannon1_K}, that is {\it not} available: smaller (larger) values of $S_\rho$ or $S_\gamma$ mean more (less) knowledge about the particle location or momentum, respectively. These two components are not independent from each other but for each bound orbital obey the relation \cite{Bialynicki1,Beckner1}
\begin{equation}\label{ShannonInequality1}
S_{t_\mathtt{n}}\geq\mathtt{d}(1+\ln\pi)
\end{equation}
with it left-hand side being a sum of the two:
\begin{equation}\label{ShannonSum1}
S_t=S_\rho+S_\gamma.
\end{equation}
Nonzero value of the right-hand side of the fundamental inequality~\eref{ShannonInequality1} demonstrates our inability to exactly know simultaneously both position and momentum of the corpuscle. For each integer $\mathtt{d}$, inequality~\eref{ShannonInequality1} is saturated by the Gaussian distribution in either space. Many tasks in data compression, quantum cryptography, entanglement witnessing, quantum metrology and other fields of quantum-information processing employ correlations between the position and momentum components of the information measures with the heavy use of the uncertainty relations \cite{Wehner1,Coles1,Toscano1,Hertz1,Wang1}.

It follows from equations~\eref{Shannon1} that the functionals $S_\rho$ and $S_\gamma$ are measured in terms of the dimensionality of the system times, respectively, the positive and negative logarithm of the length. This well-known physical ambiguity \cite{Srinivas1,Dodonov1,Olendski9,Olendski2,Olendski7,Olendski10} is a direct consequence of the fact that these expressions are extensions to the continuous distributions of the {\it unitless} Shannon entropy of a discrete case:
\begin{equation}\label{ShannonDiscrete1}
S_p=-\sum_{n=1}^Np_n\ln p_n,
\end{equation}
where $0\leq p_n\leq1$, $n=1,2,\ldots,N$, is a probability belonging to a discrete set of all $N$ (that might by finite or infinite) possible events, so that $\sum_{n=1}^Np_n=1$. Such a mapping to the continuous dependence was first proposed by C. E. Shannon himself \cite{Shannon2}. Observe that due to the different signs of the logarithms in the units of $S_\rho$ and $S_\gamma$, their sum $S_t$, obviously, is dimensionless, as it is directly seen, e.g., from inequality~\eref{ShannonInequality1}. Strictly non-negative quantity $S_p$ is confined between zero (definite event, one of the probabilities $p_n$ is equal to unity with all others disappearing) and $\ln N$ (equiprobable configuration, all happenings have the same chance to occur, $p_n=1/N$) with its continuous counterparts being able to plunge below zero \cite{Shannon2} when, for example, the parts of the corresponding density, say, $\gamma_\mathtt{n}({\bf k})$, which are larger than unity, in their contribution to the integral in equation~\eref{Shannon1_K} preponderate those segments where $\gamma_\mathtt{n}({\bf k})<1$. Also, it is very important to underline that in equations~\eref{Shannon1} the classical rule from equation~\eref{ShannonDiscrete1} is applied to the position and momentum spreadings of the \textit{pure} state that is represented by the state vector. From this point of view, $S_\rho$ and $S_\gamma$ should not be considered as 'quantum extensions' of the quantities defined for classical dependencies but rather they are 'classical quantities' applied to some quantum--related probability distributions. However, the most general quantum formulation employs the density operator $\widehat{\bm\rho}$ suitable for the description of the \textit{mixed} states. The corresponding \textit{dimensionless} von Neumann entropy
\begin{equation}\label{vonNeumann1}
\mathsf{S}\left(\widehat{\bm\rho}\right)=-{\rm Tr}\left(\widehat{\bm\rho}\ln\widehat{\bm\rho}\right)
\end{equation}
is non-negative and turns to zero for the pure states only. Since in this research we stay focused on just the pure states, below we investigate the functionals $S_\rho$ and $S_\gamma$.

Together with the Shannon entropy, which is a measure of the global behaviour of the distribution, other functionals are a popular target of scientific scrutiny too; in particular, position ${\bm\nabla}_{\bf r}$ and momentum ${\bm\nabla}_{\bf k}$ gradients that enter into the expressions of the corresponding Fisher informations
\begin{subequations}\label{Fisher1}
\begin{align}\label{Fisher1_R}
I_{\rho_\mathtt{n}}&=\int_{\mathcal{D}_\rho^{(\mathtt{d})}}\rho_\mathtt{n}({\bf r})\left|{\bm\nabla}_{\bf r}\ln\rho_\mathtt{n}({\bf r})\right|^2\!\!d{\bf r}=\int_{\mathcal{D}_\rho^{(\mathtt{d})}}\frac{\left|{\bm\nabla}_{\bf r}\rho_\mathtt{n}({\bf r})\right|^2}{\rho_\mathtt{n}({\bf r})}d{\bf r}\\
\label{Fisher1_K}
I_{\gamma_\mathtt{n}}&=\int_{\mathcal{D}_\gamma^{(\mathtt{d})}}\gamma_\mathtt{n}({\bf k})\left|{\bm\nabla}_{\bf k}\ln\gamma_\mathtt{n}({\bf k})\right|^2\!\!d{\bf k}=\int_{\mathcal{D}_\gamma^{(\mathtt{d})}}\frac{\left|{\bm\nabla}_{\bf k}\gamma_\mathtt{n}({\bf k})\right|^2}{\gamma_\mathtt{n}({\bf k})}\,d{\bf k}
\end{align}
\end{subequations}
turn the latter into the descriptors of the local performance of the associated distribution that quantify its speed of variation. Among countless applications of these measures in science, technology, engineering, medicine, finance and many other branches of human activity \cite{Frieden1}, one has to mention pivotal role of the position component in density functional theory where there exists a precise connection between it and the many-body quantum-mechanical kinetic energy functional \cite{Sears1}. This relation between information and the kinetic energy leads, in particular, to the reformulation of quantum mechanical variation principle in terms of minimization of the information. Contrary to the Shannon entropy, there is no general universal bound on the product $I_\rho I_\gamma$ but, e.g., Stam \cite{Stam1} and Cram\'{e}r-Rao \cite{Dembo1} inequalities establish relations between one of the components with covariance of the corresponding or conjugate variable. Miscellaneous aspects of the different forms of the Fisher information reveal new intriguing properties of various quantum systems \cite{Olendski1,Olendski9,Olendski2,Olendski7,Olendski10,Olendski11,Olendski3,GonzalezFerez3,Mohamed2,Mohamed1,Mohamed3}.

Position and wave vector parts of another measure considered in the present research quantitatively describe deviations of the distributions $\rho_\mathtt{n}({\bf r})$ and $\gamma_\mathtt{n}({\bf k})$ from their uniform counterparts:
\begin{subequations}\label{Onicescu1}
\begin{align}\label{Onicescu1_R}
O_{\rho_\mathtt{n}}&=\int_{\mathcal{D}_\rho^{(\mathtt{d})}}\rho_\mathtt{n}^2({\bf r})d{\bf r}\\
\label{Onicescu1_K}
O_{\gamma_\mathtt{n}}&=\int_{\mathcal{D}_\gamma^{(\mathtt{d})}}\gamma_\mathtt{n}^2({\bf k})d{\bf k}.
\end{align}
\end{subequations} 
In the literature \cite{Olendski1,Olendski9,Olendski2,Olendski7,Olendski10,Chatzisavvas1,GonzalezFerez4,Dehesa1,Ghosal1,Olendski11,Olendski3,Ou1}, functionals from equations~\eref{Onicescu1} are frequently called Onicescu energies after the Romanian mathematician who in 1966 considered their discrete counterpart \cite{Onicescu1}:
\begin{equation}\label{OnicescuDiscrete1}
O_p=\sum_{n=1}^Np_n^2.
\end{equation}
However, chronologically, the sum from equation~\eref{OnicescuDiscrete1} or its complement $1-O_p$ have been quite intensively used since at least more than half a century before O. Onicescu with a brief history described in references~\cite{Ellerman1,Good1}. Accordingly, below we follow I. J. Good advice that "it is unjust to associate" $O_p$ "with any one person" \cite{Good1} and will refer to $O_{\rho,\gamma}$ as informational energy (as Onicescu originally did \cite{Onicescu1}) or disequilibrium. Their product apparently is bounded from above according to \cite{Ghosal1}:
\begin{equation}\label{OnicescuInequality1}
O_\rho O_\gamma\leq\frac{1}{(2\pi)^\mathtt{d}}
\end{equation}
with the equality sign being achieved for the harmonic oscillator (HO) lowest level only when in either space the corresponding waveform is just a Gaussian. However, inequality~\eref{OnicescuInequality1} is not universal; namely, as will be shown below, it is violated for the Neumann QD in the magnetic field.

For the QD with its strong spatial confinement, the surface properties become an essential factor drastically influencing its characteristics. Two simplest BCs describing an interaction of the single charged particle in the nanostructure with the outer environment are the Dirichlet,
\begin{subequations}\label{BC1}
\begin{align}\label{BC1_Dirichlet}
\left.\Psi^{(D)}({\bf r})\right|_{\cal S}=0,
\intertext{and Neumann,}\label{BC1_Neumann}
\left.{\bf n}{\bm\nabla}_{\bf r}\Psi^{(N)}({\bf r})\right|_{\cal S}=0,
\end{align}
\end{subequations}
requirements with $\cal S$ being an enclosing interface and $\bf n$ its unit inward normal. Physically, the zeroing of the position waveform at the boundary by the Dirichlet BC makes it a repulsive surface that pushes the particle away from it into the QD interior whereas the vanishing gradient at $\cal S$, equation~\eref{BC1_Neumann}, turns it into the attractive shell at which the local or global maximum of $\rho({\bf r})$ is located. This difference causes, among others, a distinction between above-mentioned measures of the Dirichlet and Neumann QDs; in particular, the sum $S_t$ for each level is always smaller for the former geometry as compared to the corresponding orbital of the Neumann dot \cite{Olendski1} what means, according to the physical interpretation of the Shannon entropy discussed above, the larger amount of total information provided by the Dirichlet configuration. The previous statement holds true for the momentum component $S_\gamma$  too and both of them were confirmed for the higher dimensions $\mathtt{d}\geq3$ \cite{Olendski2}. The difference between the two types of the BCs decreases for the higher-energy levels as they are less affected by the confinement.

Present research is devoted to the comparative analysis of the uniform perpendicular magnetic field $\bf B$ influence on the quantum-information measures of the QDs with these two BC types. Before discussing our results, we briefly provide a summary of the  previous findings on the effect of the intensity $B$ on these quantities for different nanostructures. Investigation of the magnetic field impact on the quantum-information measures has already its own history; in particular, a study of the dynamics of some excited states of hydrogen in the presence of parallel magnetic and electric fields pointed out the role of the position Shannon entropy as an indicator or predictor of the avoided crossings in the spectrum and it was found that $S_\rho$ manifests the informational exchange of the involved states as the magnetic field strength is varied \cite{GonzalezFerez1,GonzalezFerez2}. Very similar conclusions were drawn for the position Fisher information too \cite{GonzalezFerez3}. Scaling properties of composite information measures and shape
complexity for the same system were studied too \cite{GonzalezFerez4}. Properties of all three measures were thoroughly analyzed for the Volcano-type Aharonov-Bohm ring in the uniform magnetic field \cite{Olendski3} and, among other findings, it was shown that the sum $S_\rho+S_\gamma$ and products $I_\rho I_\gamma$ and $O_\rho O_\gamma$ are $B$-independent quantities what means, in particular, that an amount of the total information that is available simultaneously about position and motion of the nanoparticle can not be altered by the uniform magnetic fields. Research was expanded to the R\'{e}nyi and Tsallis entropies \cite{Olendski4} which are one-parameter generalizations of their Shannon counterpart. Functionals $S_\rho$ and $S_\gamma$ of a spinless non--Hermitian particle in the presence of a magnetic field were studied \cite{Lima1}. The measures' dependencies on the uniform field were calculated for the particle moving in the Yukawa-type potential \cite{Edet1,Edet2}. The information content of one--electron bulk and edge states of semiconductor HgTe-CdTe quantum wells in magnetic field was considered in the inverted regime \cite{Giovenale1}. Influence of the magnetic intensity $B$, Aharonov-Bohm flux and the topological defect on the position and momentum Shannon functionals was theoretically analyzed for the ring with Yukawa interaction in curved space with disclination \cite{Edet3}. Concerning our geometry, an attempt has been made to calculate the influence of the homogeneous field $\bf B$ on the Shannon entropies of the Dirichlet dot \cite{Cruz1}.

The first step in our discussion is a derivation and investigation of the transcendental equations for finding eigen energies of the particle dwelling in the hard-wall either Dirichlet or Neumann QD and subject to the field. Even though this part of the problem was previously tackled by several researchers, we point out some new properties not discussed before; in particular, application of the perturbation theory in the regime of both weak and strong fields leads to the simple expressions that shed lights on the similarities and differences between the two BCs. Physically, the fundamental distinction stems from the fact that the Dirichlet surface acts as a {\it repulsive} border with its Neumann counterpart {\it attracting} the particle. Thus, the former helps the increasing magnetic intensity to push the corpuscle to the QD center whereas the latter BC opposes the field in this endeavour. Position waveforms $\Psi$ evolution with the field is analyzed as well and it is shown how the growing $B$ decreases the difference between the Dirichlet and Neumann dependencies. The same is performed for the momentum functions $\Phi$  too that are evaluated as a Fourier transform of their position fellows. Knowledge of both $\Psi$ and $\Phi$ opens the way for the efficient calculation of all three introduced above measures in position and wave vector representations; in particular, the aspects of convergence of the Shannon entropies with the increasing field that transforms the fundamental inequality~\eref{ShannonInequality1} into the strict identity are analyzed in detail. As a by-product, we also derive expressions of all three measures for the parabolic QD and compare them with the hard-wall confinement.

We proceed as follows. Sec.~\ref{Sec_Energy} addresses the energies and associated functions in either space with the emphasis on their evolution with the field. Sec.~\ref{Sec_Measures} is devoted to the analysis of the quantum-information measures and magnetic field influence on them with separate subsections on the Shannon entropy, Fisher information and informational energy. The discussion is wrapped up in Sec.~\ref{Sec_Conclusions} by some conclusions.

\section{Energy spectrum and position and momentum waveforms}\label{Sec_Energy}
A charged quantum particle (for definiteness, we will talk about the electron, $q=-|e|$) is free to move, $V({\bf r})\equiv 0$, inside the flat disc of radius $a$ of the $x-y$ plane whose circumference supports either Dirichlet, equation~\eref{BC1_Dirichlet}, or Neumann, equation~\eref{BC1_Neumann}, BC. In addition, magnetic field ${\bf B}\equiv B\hat{k}$ is applied normally to the structure, $\hat{k}$ is a unit vector along the $z$-axis. Geometry of the system dictates a choice of the polar coordinates, ${\bf r}\equiv(r,\varphi_{\bf r})$, $0\leq r\leq a$, $0\leq\varphi_{\bf r}<2\pi$, with the origin at the center of the circle, as the most natural and convenient one. Then, in a symmetric gauge for the vector potential ${\bf A}=\frac{1}{2}{\bf B}\times{\bf r}$ one has $A_r=0$, $A_{\varphi_{\bf r}}=\frac{1}{2}Br$, what shapes the Schr\"{o}dinger equation for the 2D waveform $\Psi(r,\varphi_{\bf r})$ into the following form:
\begin{equation}\label{Schrodinger2}
-\frac{\hbar^2}{2M}\left[\frac{1}{r}\frac{\partial}{\partial r}\left(r\frac{\partial\Psi}{\partial r}\right)+\frac{1}{r^2}\frac{\partial^2\Psi}{\partial\varphi_{\bf r}^2}\right]-i\frac{\hbar\omega_c}{2}\frac{\partial\Psi}{\partial\varphi_{\bf r}}+\frac{M\omega_c^2}{8}r^2\Psi=E\Psi,
\end{equation}
$\omega_c=\frac{|e|B}{M}$ is a cyclotron frequency. Separation of variables
\begin{equation}\label{Separation1}
\Psi_{nm}(r,\varphi_{\bf r})=\frac{e^{im\varphi_{\bf r}}}{(2\pi)^{1/2}}\,R_{nm}(r)
\end{equation}
with $n$ and $m$ being principal and angular quantum numbers, respectively, $n=0,1,2,\ldots$, $m=0,\pm1,\pm2,\ldots$, yields the following differential equation for the radial dependence $R_{nm}(r)$:
\begin{equation}\label{RadialEquation1}
\frac{\hbar^2}{2M}\left(\frac{d^2}{dr^2}+\frac{1}{r}\frac{d}{dr}-\frac{m^2}{r^2}\right)\!R+\left(E-\frac{M\omega_c^2}{8}r^2-m\frac{\hbar\omega_c}{2}\right)\!R=0.
\end{equation}
Solution that stays finite at the origin is represented by the Kummer confluent hypergeometric function $M(b,c,\eta)$ \cite{Abramowitz1,Mathews1}:
\begin{subequations}\label{RadialSolutionNormalization1}
\begin{align}\label{RadialSolution1}
R_{nm}(r)&=N_{nm}\exp\!\left(-\frac{1}{4}\frac{r^2}{l_B^2}\right)\left(\frac{1}{2}\frac{r^2}{l_B^2}\right)^{\!\!|m|/2}\!\!\!\!M\!\left(\frac{m+|m|+1}{2}-\frac{E_{nm}}{\hbar\omega_c},|m|+1,\frac{1}{2}\frac{r^2}{l_B^2}\right),
\intertext{$l_B=[\hbar/(|e|B)]^{1/2}$ is a magnetic length, and constant $N_{nm}$ is found from the normalization condition:}
\label{NormalizationConstant1}
N_{nm}&=\left[\int_0^a\!\!\exp\!\left(-\frac{1}{2}\frac{r^2}{l_B^2}\right)\left(\frac{1}{2}\frac{r^2}{l_B^2}\right)^{\!\!|m|}\!\!\!\!M^{\,2}\!\left(\frac{m+|m|+1}{2}-\frac{E_{nm}}{\hbar\omega_c},|m|+1,\frac{1}{2}\frac{r^2}{l_B^2}\right)rdr\right]^{-1/2}.
\end{align}
\end{subequations}
For the Dirichlet case, an equivalent analytical expression for $N_{nm}$ contains a derivative of the energy eigenvalue with respect to the radius of the circle \cite{dePrunele1}. At this stage, in order to make our results more universal and compact, it is convenient to introduce dimensionless units. For doing this, let us point out that the system is characterized by the two types of quantization: electric and magnetic one with the former (latter) being dominant at the small (huge) $B$. Accordingly, there are two length scales: dot radius $a$ and magnetic length $l_B$. In the same way, there are two reference energies: field-free ground-state energy of the 1D Dirichlet quantum well $\pi^2\hbar^2/(2Ma^2)$ (for electric quantization) and $\hbar\omega_c$ (when magnetic interaction is dominant). Thus, below, the overlined symbols refer to the quantities normalized in electric units: $\overline{r}=r/a$, $\overline{k}=ak$, $\overline{E}=E\left/\left[\pi^2\hbar^2/(2Ma^2)\right]\right.$ etc. whereas doubly overlined designations imply their regular counterparts expressed in magnetic denominations: $\overline{\overline{r}}=r/l_B$, $\overline{\overline{E}}=E/(\hbar\omega_c)$, and so on. Under this convention, the magnetic field is measured relative to $\hbar/(|e|a^2)$, such that $\overline{B}=B/[\hbar/(|e|a^2)]$. In these units, two last equations read:
%
\begin{align}\tag{19a$'$}\label{eq:19a'}
R_{nm}(r)&=N_{nm}\exp\!\left(-\frac{1}{4}\overline{B}\,\overline{r}^{\,2}\right)\left(\frac{1}{2}\overline{B}\,\overline{r}^{\,2}\right)^{\!\!|m|/2}\!\!\!\!M\!\left(\frac{m+|m|+1}{2}-\frac{\pi^2}{2}\frac{\overline{E}_{nm}}{\overline{B}},|m|+1,\frac{1}{2}\overline{B}\,\overline{r}^{\,2}\right)\\
\tag{19b$'$}\label{eq:19b'}
N_{nm}&=\frac{1}{a}\left[\int_0^1\!\!\exp\!\left(-\frac{1}{2}\overline{B}\,\overline{r}^{\,2}\right)\left(\frac{1}{2}\overline{B}\,\overline{r}^{\,2}\right)^{\!\!|m|}\!\!\!\!M^{\,2}\!\left(\frac{m+|m|+1}{2}-\frac{\pi^2}{2}\frac{\overline{E}_{nm}}{\overline{B}},|m|+1,\frac{1}{2}\overline{B}\,\overline{r}^{\,2}\right)\overline{r}d\overline{r}\right]^{-1/2}.
\end{align}
%
For our geometry, BCs from equations~\eref{BC1} degenerate to
\begin{subequations}\label{BC2}
\begin{align}\label{BC2_Dirichlet}
R_{nm}^{(D)}(a)&=0\\
\label{BC2_Neumann}
\left.\frac{d}{dr}R_{nm}^{(N)}(r)\right|_{r=a}&=0,
\end{align}
\end{subequations}
what leads to the transcendental relations determining the corresponding Dirichlet $E_{nm}^{(D)}(B)$ \cite{Geerinckx1,Constantinou1} and Neumann $E_{nm}^{(N)}(B)$ \cite{Constantinou2} energy dependencies:
\begin{subequations}\label{Energies1}
\begin{align}\label{Energies1_D}
M\!\!\left(\frac{m+|m|+1}{2}-\frac{\pi^2}{2}\frac{\overline{E}_{nm}^{\,(D)}}{\overline{B}},|m|+1,\frac{1}{2}\overline{B}\right)=0\\
\left(\frac{|m|}{\overline{B}}-\frac{1}{2}\right)\!M\!\!\left(\frac{m+|m|+1}{2}-\frac{\pi^2}{2}\frac{\overline{E}_{nm}^{\,(N)}}{\overline{B}},|m|+1,\frac{1}{2}\overline{B}\right)\nonumber\\
\label{Energies1_N}
+M'\!\left(\frac{m+|m|+1}{2}-\frac{\pi^2}{2}\frac{\overline{E}_{nm}^{\,(N)}}{\overline{B}},|m|+1,\frac{1}{2}\overline{B}\right)=0.
\end{align}
\end{subequations}
\begin{figure}
\centering
\includegraphics[width=\columnwidth]{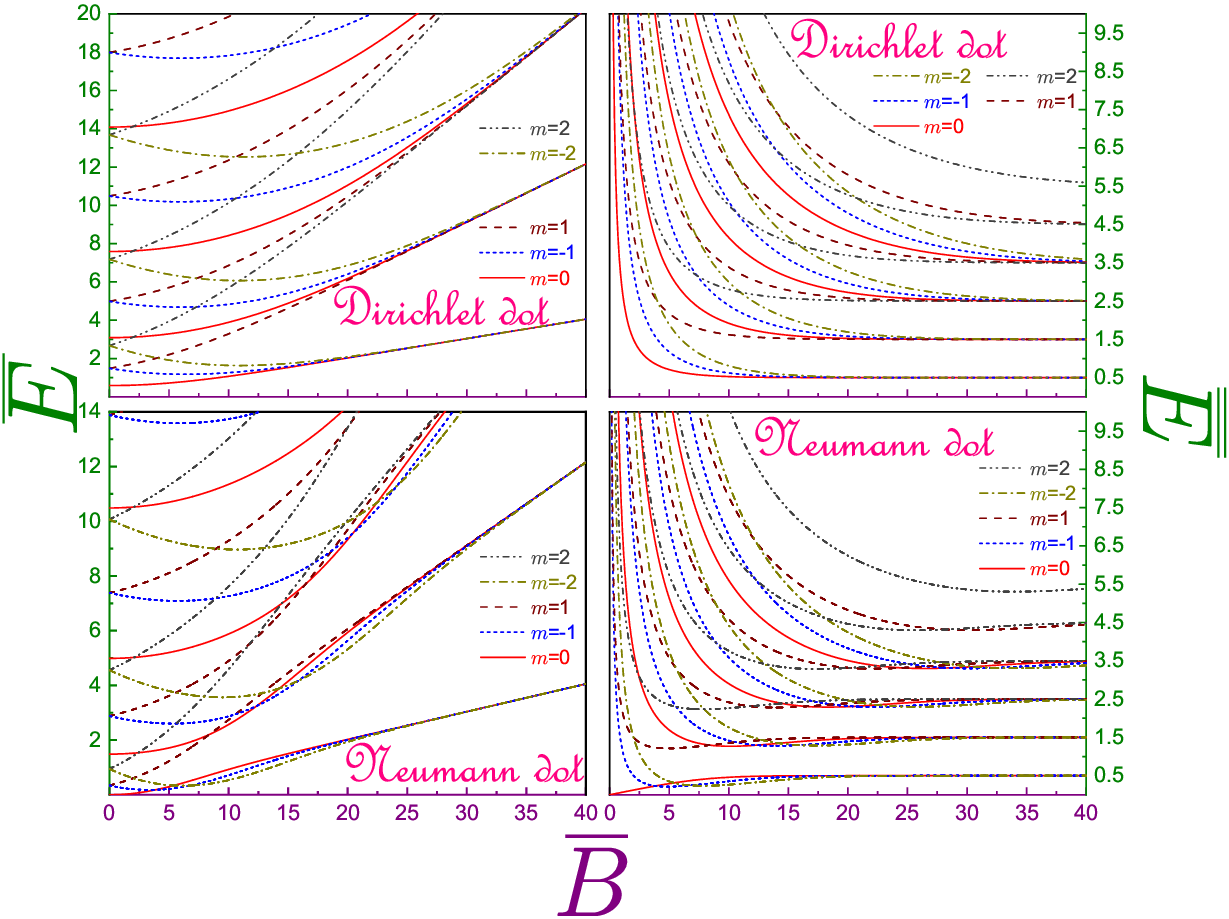}
\caption{\label{Fig_Energies1}
Dimensionless energies $\overline{E}_{nm}$ (left panels) and $\overline{\overline{E}}_{nm}$ (right windows) of the Dirichlet (upper subplots) and Neumann (lower figures) QD as functions of the normalized magnetic field $\overline{B}$ for $0\leq n\leq3$ and $0\leq|m|\leq2$. Rotationally symmetric, $m=0$, orbitals are depicted by the solid lines, dotted curves are for the states with $m=-1$, dashed ones -- for  $m=1$,  dash-dotted lines are for the levels with  $m=-2$ and dash-dot-dotted dependencies show $m=2$ orbitals. In each set, higher-lying lines denote larger index $n$. Note different vertical ranges of the left figures.}
\end{figure}

Figure~\ref{Fig_Energies1} exhibits both Dirichlet (upper plots) and Neumann (lower panels) spectra where, in addition to the functions $\overline{E}_{nm}\!\left(\overline{B}\right)$, the dependencies $\overline{\overline{E}}_{nm}\!\left(\overline{B}\right)$ are also shown for better clarity. Despite the fact that the former curves are quite well known \cite{Cruz1,Geerinckx1,Constantinou1,Constantinou2,Robnik1,Lent1,Schult1,Avishai1,Zawadzki1,Tanaka1,Spehner1}, they are included here for completeness since a comparative analysis of the two geometries will point out to some important properties not discussed before. First, for either BC, the zero-field degeneracy of the $m\neq0$ states is lifted by the intensity $B$: as it follows from equations~\eref{Energies1}, the difference $E_{nm}-E_{n,-m}$ is just the $m$ multiple of the cyclotron energy:
\begin{equation}\label{EnergyDeifference1}
E_{nm}^{(D,N)}-E_{n,-m}^{(D,N)}=m\hbar\omega_c.
\end{equation}
Next, the small magnetic fields influence differently the $m=0$ and $m\neq0$ orbitals: an application of the standard perturbation theory in the regime $\overline{B}\ll1$ with the use of the properties of the integrals of the Bessel functions \cite{Prudnikov1} reveals that the energies of the former levels increase quadratically with the field whereas the latter depend linearly on diminutive $B$:
\begin{subequations}\label{Perturbation1}
\begin{align}\label{Perturbation1_Dirichlet}
E_{nm}^{(D)}(B)&=\frac{\hbar^2}{2Ma^2}j_{|m|,n+1}^2+\left\{\begin{array}{cc}
\frac{1}{24}\left(1-\frac{2}{j_{0,n+1}^2}\right)\frac{|e|^2a^2}{M}B^2,& m=0\\
\frac{m}{2}\hbar\omega_c,& m\neq0
\end{array}
\right.\\
\label{Perturbation1_Neumann}
E_{nm}^{(N)}(B)&=\left\{\begin{array}{cc}
\left\{\begin{array}{cc}
\frac{1}{16}\frac{|e|^2a^2}{M}B^2,&n=0\\
\frac{\hbar^2}{2Ma^2}\!\left(j_{0,n+1}'\right)^{\!2}+\frac{1}{24}\frac{|e|^2a^2}{M}B^2,& n\geq1
\end{array}
\right\},&m=0\\
\frac{\hbar^2}{2Ma^2}\!\left(\!j_{|m|,n+1}'\!\right)^{\!2}+\frac{m}{2}\hbar\omega_c,& m\neq0,
\end{array}
\right.
\end{align}
\end{subequations}
with $j_{\nu l}$ and $j_{\nu l}'$ being $l$th root, $l=1,2,\ldots$, of the $\nu$th Bessel function of the first kind $J_\nu(x)$ and its derivative $J_\nu'(x)$, respectively \cite{Abramowitz1}. These relations show that the zero-field degeneracy of the $m\neq0$ levels is destroyed in such a way that the states with positive (negative) angular index become,  according to the equation for the magnetic moment
\begin{equation}\label{MagneticMoment1}
{\bf M}=-\frac{\partial E}{\partial{\bf B}},
\end{equation}
diamagnetics (paramagnetics). Applied field breaks time reversal symmetry of the system forcing the electrons with $m>0$ ($m<0$) to increase (decrease) their energies. The $m=0$ orbitals exhibit diamagnetic behaviour too with the magnitude of the magnetic moment being in this regime a linear function of the small intensity $B$. With the further increase of the field, the quadratic term in the second parenthesis of the left-hand side of equation~\eref{RadialEquation1} starts to dominate over its linear counterpart what for both BCs results in the minimum of the functions $\overline{E}_{nm}\left(\overline{B}\right)$ that, for the different negative $m$, is achieved at the different intensities $B$. Making fields even stronger turns all orbitals into the diamagnetic contributors that smoothly transform at $B\rightarrow\infty$ into the Landau levels with their energies $E^{({\bf B})}_{nm}$ and wave functions $R_{nm}^{({\bf B})}(r)$ being \cite{Kim1}:
\begin{subequations}\label{UniformField1}
\begin{align}\label{UniformField1_Energy}
E_{nm}^{({\bf B})}&=\hbar\omega_c\!\left(n+\frac{m+|m|+1}{2}\right)\\
\label{UniformField1_PositionFunction}
R_{nm}^{({\bf B}	)}(r)&=\frac{1}{l_B}\!\left[\frac{n!}{(n+|m|)!}\right]^{1/2}\!\!\exp\!\left(-\frac{1}{4}\frac{r^2}{l_B^2}\right)\!\left(\frac{1}{2}\frac{r^2}{l_B^2}\right)^{\!\!|m|/2}\!L_n^{|m|}\!\left(\frac{1}{2}\frac{r^2}{l_B^2}\right),
\end{align}
\end{subequations}
$L_n^{(\alpha)}(x)$ is a generalized Laguerre polynomial \cite{Abramowitz1}. A crucial difference between the two BCs is the fact that for the Neumann QD, contrary to the Dirichlet case, the energies of the levels with the same $n$ and adjacent non-positive $m$ during the process of 'Landau condensation' \cite{Robnik1} cross each other at the increasing field with the intersecting point for the larger $|m|$ being achieved at the stronger intensities; for example, for the ground band, $n=0$, the $m=0$ and $m=-1$ orbitals degenerate at $\overline{B}\approx3.848$ when both energies are equal to $\overline{E}=0.180$ or $\overline{\overline{E}}=0.231$; for the $m=-1$ and $m=-2$ levels these numbers are $\overline{B}\approx6.785$, $\overline{E}=0.338$ and $\overline{\overline{E}}=0.247$, etc. To explain this phenomenon, one has to mention first that the growth of the magnetic field pushes the electron closer to the QD centre, i.e., its influence coincides with that from the Dirichlet circle, which repels the particle just in the same direction, whereas the Neumann wall tries to pull the electron out of the middle of the dot, as discussed in the Introduction, with its attraction additionally lowering the energy that already decreases by the amount of the linear item in eqiation~\eref{RadialEquation1}. Thus, the Dirichlet BC and the field act in unison whereas the Neumann interface competes with magnetic intensity. Since the larger $|m|$ means a further shift from the origin and stronger localization at the edge \cite{Olendski1}, higher fields are needed for them to overcome the Neumann attraction; thus, there is a range of the intensities $B$ inside which the energy of the state with the negative $m$ already increases whereas that of the $m-1$ orbital still decreases. These crossings are a common feature of the quantum systems in the magnetic fields where the charged particle experiences an obstacle in reaching the centre, such as, for example, Volcano-type quantum ring \cite{Olendski5} or two electrons in parabolic QD \cite{Merkt1,Wagner1,DeGroote1,Olendski6} where in the relative motion a Coulomb interaction pushes the particles away from each other.

As stated above, at the strong fields the levels collapse into the regular Landau orbitals. This Landau condensation is clearly seen in the right panels. In this regime, the dot surface presents a minor disturbance to the orbitals described by equations~\eref{UniformField1}. Applying perturbation theory that was developed for similar situations by Brillouin \cite{Briullouin1} and Froehlich \cite{Froehlich1}, one finds that the energy corrections $\Delta E$ to the uniform Landau levels, equation~\eref{UniformField1_Energy}, read:
\begin{equation}\label{Correction1}
\Delta E^{(D,N)}=\mp\frac{\hbar^2a}{2M}R_{nm}^{({\bf B})}(a){R_{nm}^{({\bf B})}}'(a).
\end{equation}
At the strong fields, $a\gg l_B$, the radial dependence at $r=a$ smoothly fades to zero with the sign of the derivative ${R_{nm}^{({\bf B})}}'(r)$ at this point being opposite to that of $R_{nm}^{({\bf B})}(a)$ what means that the Dirichlet (Neumann) energies approach Landau levels from above (below) what is easily explained by the repulsive (attractive) role played by the corresponding surface. Equation~\eref{Correction1}, which is a first-order perturbation correction, implies that the two BC energies are equally spaced from the corresponding Landau state. Exact calculations reveal that the Dirichlet one is slightly closer to $E_{nm}^{({\bf B})}$ with this minuscule mismatch quickly waning at the growing $B$. In addition, the states of the smaller $|m|$ are closer to the Landau energies since, as mentioned above, the corresponding wave functions are localized further from the dot perimeter.

\begin{figure}
\centering
\includegraphics[width=\columnwidth]{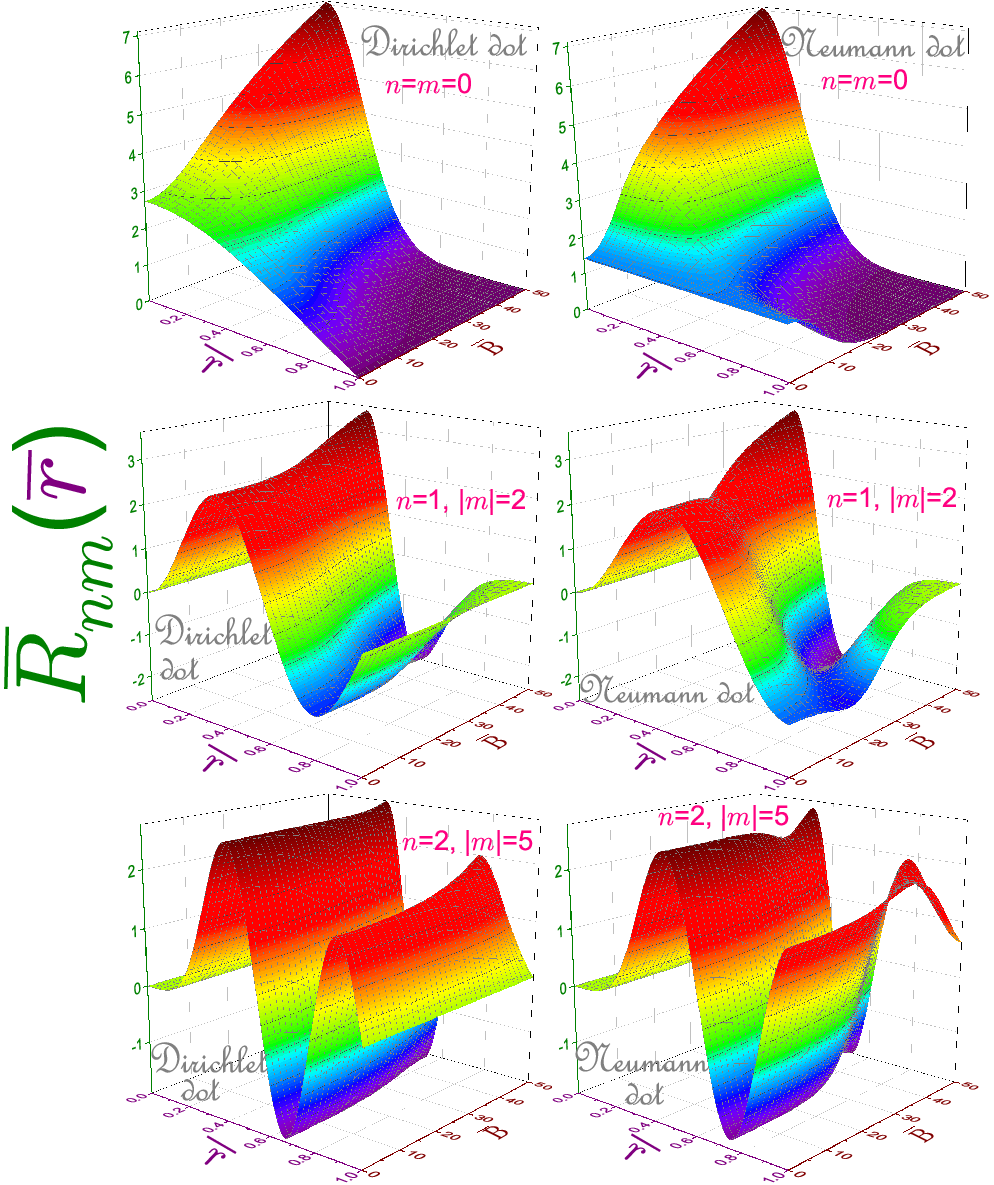}
\caption{\label{Fig_PositionFunctions}
Position waveforms $\overline{R}_{nm}(\overline{r})$ evolution with the normalized magnetic field $\overline{B}$: upper windows exhibit the states with $n=m=0$, middle rows - $n=1$, $|m|=2$ levels and lower subplots show $n=2$, $|m|=5$ orbitals whereas left (right) panels are for the Dirichlet (Neumann) BC. Note different vertical ranges for each horizontal row of windows.}
\end{figure}

Knowledge of the energy spectrum completely determines, according to equations~\eref{RadialSolutionNormalization1} or (19$'$), radial waveforms which at either BC obey the orthonormalization:
\begin{equation}\label{OrthoNormalityR1}
\int_0^aR_{nm}(r)R_{n'm}(r)rdr=\delta_{nn'},
\end{equation}
$n'=0,1,2,\ldots$. Figure~\ref{Fig_PositionFunctions} exhibits position functions $\overline{R}_{nm}(\overline{r})$ evolution with the field for the three different levels and both BCs. For the $n=m=0$ orbital, the Neumann $B=0$ function is just a constant with its Dirichlet counterpart monotonically decreasing from its maximum at the dot centre to the vanishing value at the rim \cite{Olendski1}, see upper windows. The increase of the magnetic intensity pushes the distributions closer to the origin subduing in this way the influence of the border what results in the decrease of the difference between the two wave functions and at the huge fields they tend to become identical. Qualitatively, the same behaviour is characteristic for all other levels too with the dissimilarity at the larger $n$ and $|m|$ persisting at the higher $B$. Let us also note that, for either BC and at arbitrary magnetic fields, the rotationally symmetric states, $m=0$, are the only ones that allow the particle to reach the center,
\begin{equation}\label{RadialPositionZero1}
R_{nm}(0)=N_{nm}\delta_{m,0},
\end{equation}
as it directly follows from equations~\eref{RadialSolution1} or \eref{eq:19a'}. 

\begin{figure}
\centering
\includegraphics[width=\columnwidth]{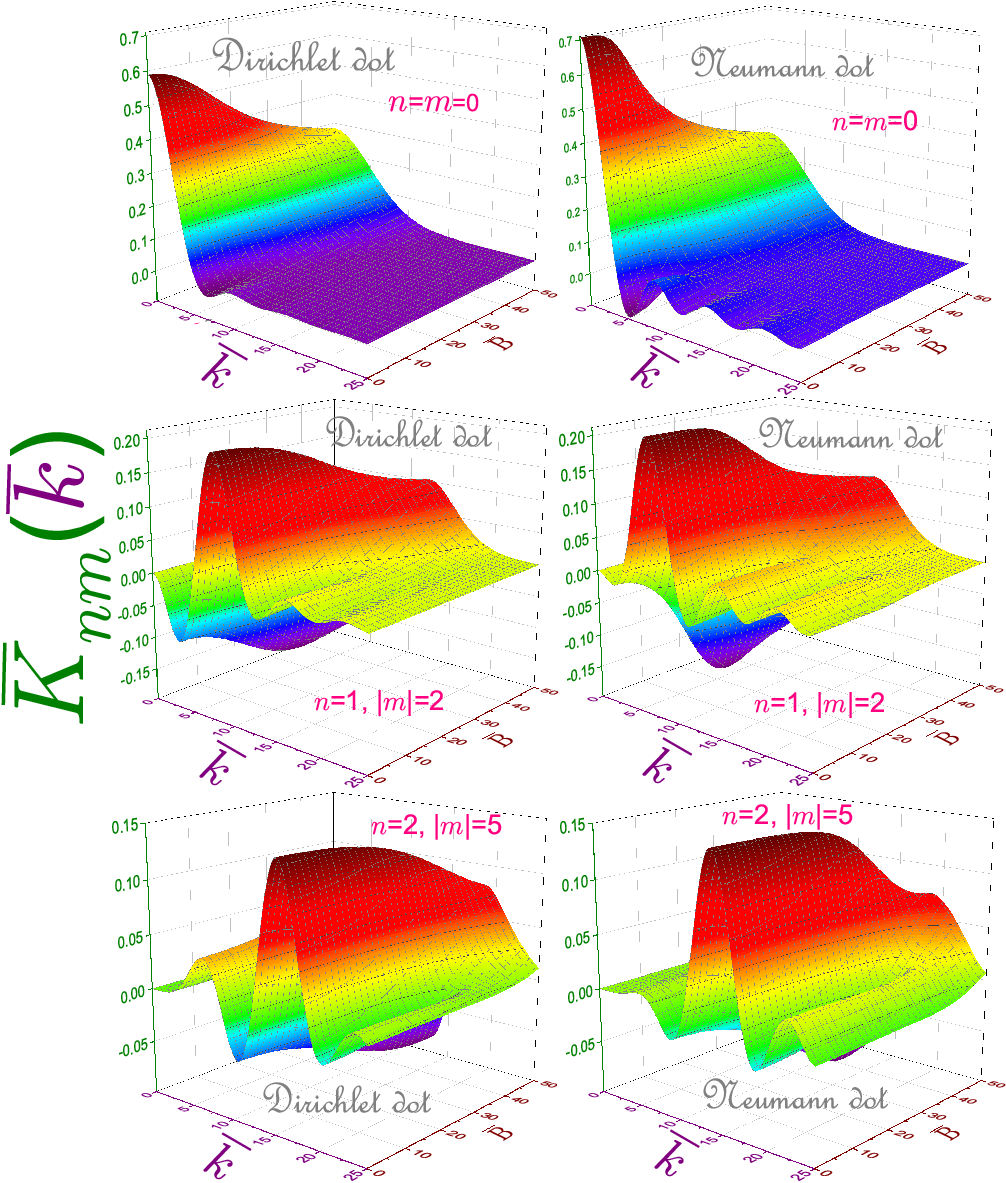}
\caption{\label{Fig_MomentumFunctions}
The same as in figure~\ref{Fig_PositionFunctions} but for the momentum functions $\overline{K}_{nm}\left(\overline{k}\right)$.} 
\end{figure}

Having determined position dependencies, it is possible to calculate, according to equation~\eref{Fourier1_1}, their momentum counterparts:
\begin{equation}
\Phi_{nm}(k,\varphi_{\bf k})=\frac{1}{2\pi}\int_0^adrrR_{nm}(r)\int_0^{2\pi}d\varphi_{\bf r}e^{i[m\varphi_{\bf r}-kr\cos(\varphi_{\bf r}-\varphi_{\bf k})]}.
\end{equation}
Similar to other symmetric structures \cite{Olendski1,Olendski3,Olendski4}, azimuthal integration is carried out analytically what allows to represent the function $\Phi$ as a product of the angular and radial dependencies too:
\begin{equation}\label{MomentumFunction1}
\Phi_{nm}\!\left(k,\varphi_{\bf k}\right)=\frac{(-i)^m}{(2\pi)^{1/2}}e^{im\varphi_{\bf k}}K_{nm}(k)
\end{equation}
with
\begin{equation}\label{MomentumRadialFunction1}
K_{nm}(k)=\int_0^arR_{nm}(r)J_{|m|}(kr)dr.
\end{equation}
Obviously, position orthonormalization from equation~\eref{OrthoNormalityR1} is inherited by the function $K_{nm}(k)$:
\begin{equation}\label{OrthonormalityK1}
\int_0^\infty K_{nm}(k)K_{n'm}(k)kdk=\delta_{nn'}.
\end{equation}
At the nonzero $B$, contrary to the field-free configuration \cite{Olendski1}, analytic expressions for the type of the integrals from the right-hand side of equation~\eref{MomentumRadialFunction1} are missing in the known literature \cite{Abramowitz1,Prudnikov1,Prudnikov2,Brychkov1,Gradshteyn1}; accordingly, their direct numerical quadrature was used. Figure~\ref{Fig_MomentumFunctions} shows momentum dependencies for the same orbitals as in figure~\ref{Fig_PositionFunctions}. Zero-field Dirichlet functions at the large wave vectors, $ak\gg1$, behave like $J_{|m|}(ak)/(ak)^2$ with their Neumann fellows obeying the $J_{|m|}'(ak)/(ak)$ rule \cite{Olendski1} what means that the fading oscillations decrease faster for the former geometry. This is clearly seen from a comparison of the upper panels. Turning on the magnetic intensity levels the oscillations and simultaneously decreases their amplitudes and diminishes the difference between the two functions, which at the huge $B$ tend asymptotically to their non-disturbed uniform field counterpart $K_{nm}^{({\bf B})}(k)$. The latter can be calculated analytically; namely, substituting into the expression
\begin{align}\tag{31$'$}\label{eq:31'}
K_{nm}^{({\bf B})}(k)&=\int_0^\infty rR_{nm}^{({\bf B})}(r)J_{|m|}(kr)dr,
\intertext{the radial dependence from equation~\eref{UniformField1_PositionFunction}, one gets:}\tag{31$''$}\label{eq:31''}
K_{nm}^{({\bf B})}(k)&=l_B\!\left[\frac{n!}{(n+|m|)!}\right]^{1/2}\!\!\int_0^\infty\!\!\exp\!\left(-\frac{1}{4}\overline{\overline{r}}^{\,2}\right)\!\left(\frac{1}{2}\overline{\overline{r}}^{\,2}\right)^{\!\!|m|/2}\!\!L_n^{|m|}\!\left(\frac{1}{2}\overline{\overline{r}}^{\,2}\right)\!J_{|m|}\!\left(\overline{\overline{k}}\overline{\overline{r}}\right)\overline{\overline{r}}d\overline{\overline{r}}.
\end{align}
Explicit evaluation yields \cite{Prudnikov1}:
\begin{align}\tag{25c}\label{eq:25c}
K_{nm}^{({\bf B})}(k)&=l_B2^{\frac{|m|}{2}+1}\left[\frac{n!}{(n+|m|)!}\right]^{1/2}e^{-(l_Bk)^2}(l_Bk)^{|m|}L_n^{|m|}\!\left(2l_B^2k^2\right);
\intertext{in particular,}
\tag{25c$'$}\label{eq:25c'}
K_{0m}^{({\bf B})}(k)&=l_B\frac{2^{\frac{|m|}{2}+1}}{(|m|!)^{1/2}}(l_Bk)^{\,|m|}e^{-(l_Bk)^2}
\intertext{and}
\tag{25c$''$}\label{eq:25c''}
K_{00}^{({\bf B})}(k)&=2l_Be^{-(l_Bk)^2}.
\end{align}
It is elementary to check that the functions from equation~\eref{eq:25c} do satisfy orthonormalization, equation~\eqref{OrthonormalityK1}. Similar to the position space, at the intensity $B$ growing the approach to these functions for the levels with higher $n$ and $|m|$ takes place at the stronger fields. Let us also point out that the position feature, equation~\eref{RadialPositionZero1}, holds for the momentum waveforms too:
\begin{equation}\label{RadialMomentumZero1}
K_{nm}(0)=\delta_{m,0}\int_0^arR_{nm}(r)dr,
\end{equation}
as it directly follows from equation~\eref{MomentumRadialFunction1} and properties of the Bessel functions \cite{Abramowitz1}. Equation \eref{RadialMomentumZero1} manifests that the rotationally symmetric states, $m=0$, are the only ones that allow the particle to have zero momentum.

\section{Quantum-information measures}\label{Sec_Measures}
We start this section from the brief analysis of the units in which Shannon entropies, Fisher informations and disequilibria are measured \cite{Olendski2,Olendski7}; namely, each spatially confined structure has some characteristic length $L$ and then, as it directly follows from equations~\eref{Shannon1} and discussed in Introduction, the functionals $S$ are expressed in terms of the positive (position component) or negative (momentum part) logarithm of $L$ times dimensionality together with the unitless $L$-independent quantity. If the system possesses more than one length, then either of them can be used as a logarithm argument with the dimensionless component being a function of the ratios of the remaining distances to that entering the logarithm. Since for our geometry there are two such ranges $a$ and $l_B$, one has:
\begin{subequations}\label{ShannonUnits1}
\begin{align}\label{ShannonUnits1_R}
S_\rho&=2\ln a+\overline{S}_\rho\!\left(\overline{B}\right)=2\ln l_B+\overline{\overline{S}}_\rho\!\left(\overline{B}\right)\\
\label{ShannonUnits1_K}
S_\gamma&=-2\ln a+\overline{S}_\gamma\!\left(\overline{B}\right)=-2\ln l_B+\overline{\overline{S}}_\gamma\!\left(\overline{B}\right).
\intertext{Accordingly, below we will study the $\overline{S}$ and $\overline{\overline{S}}$ variations with the field $\overline{B}$ what makes this method the most general and universal one. Obviously,}\label{ShannonRelation1}
\overline{S}_{\rho,\gamma}&=\mp\ln\overline{B}+\overline{\overline{S}}_{\rho,\gamma}
\intertext{and}\label{ShannonRelation2}
S_\rho+S_\gamma&=\overline{S}_\rho+\overline{S}_\gamma=\overline{\overline{S}}_\rho+\overline{\overline{S}}_\gamma.
\end{align}
\end{subequations}
Similarly, for the Fisher information the following units are used:
\begin{subequations}\label{FisherUnits1}
\begin{align}\label{FisherUnits1_R}
I_\rho&=\frac{\overline{I}_\rho\!\left(\overline{B}\right)}{a^2}=\frac{\overline{\overline{I}}_\rho\!\left(\overline{B}\right)}{l_B^2}\\
\label{FisherUnits1_K}
I_\gamma&=a^2\overline{I}_\gamma\!\left(\overline{B}\right)=l_B^2\overline{\overline{I}}_\gamma\!\left(\overline{B}\right),
\intertext{where the dimensionless quantities $\overline{I}$ and $\overline{\overline{I}}$ are related as:}
\overline{I}_\rho&=\overline{B}\,\overline{\overline{I}}_\rho\\
\overline{I}_\gamma&=\frac{\overline{\overline{I}}_\gamma}{\overline{B}}
\intertext{with}\label{FisherRelation1}
I_\rho I_\gamma&=\overline{I}_\rho\overline{I}_\gamma=\overline{\overline{I}}_\rho\overline{\overline{I}}_\gamma.
\end{align}
\end{subequations}
Exactly the same equations hold true for the 2D informational energies.

Before discussing the Dirichlet and Neumann QD functionals, it is worthwhile to provide their expressions for the uniform magnetic field or, even more generally, for the 2D HO with frequency $\omega_0$ subject to the intensity $\bf B$ \cite{Olendski3}. Its energies and position wave functions have been known for a long time \cite{Fock1,Darwin1,Dingle1}:
\begin{subequations}\label{HOB1}
\begin{align}\label{HOB1_Energy}
E_{nm}^{(HO{\bf B})}&=\hbar\omega_{eff}(2n+|m|+1)+\frac{1}{2}m\hbar\omega_c\\
\label{HOB1_Position}
R_{nm}^{(HO{\bf B})}(r)&=\frac{1}{l_{eff}}\!\left[\frac{n!}{(n+|m|)!}\right]^{1/2}\!\!\exp\!\left(-\frac{1}{4}\frac{r^2}{l_{eff}^2}\right)\!\left(\frac{1}{2}\frac{r^2}{l_{eff}^2}\right)^{\!\!|m|/2}\!L_n^{|m|}\!\left(\frac{1}{2}\frac{r^2}{l_{eff}^2}\right)
\intertext{with $\omega_{eff}=\left(\omega_0^2+\frac{1}{4}\omega_c^2\right)^{1/2}$ and $l_{eff}=\left[\hbar/\left(2M\omega_{eff}\right)\right]^{1/2}$. Momentum dependencies are found in the same way as it was done for the uniform field only, equation~\eref{eq:25c}:}
\label{HOB1_Momentum}
K_{nm}^{(HO{\bf B})}(k)&=(-1)^nl_{eff}2^{\frac{|m|}{2}+1}\left[\frac{n!}{(n+|m|)!}\right]^{1/2}e^{-(l_{eff}k)^2}(l_{eff}k)^{|m|}L_n^{|m|}\!\left(2l_{eff}^2k^2\right).
\end{align}
\end{subequations}
Without electrostatic confinement, $\omega_0=0$, these relations transform to equations~\eref{UniformField1}, as expected, and for the field-free parabolic QD one has:
\begin{subequations}\label{HO1}
\begin{align}\label{HO1_Energy}
E_{nm}^{(HO)}&=\hbar\omega_0(2n+|m|+1)\\
\label{HO1_Position}
R_{nm}^{(HO)}(r)&=\frac{2^{1/2}}{l_0}\!\left[\frac{n!}{(n+|m|)!}\right]^{1/2}\!\!\exp\!\left(-\frac{1}{2}\frac{r^2}{l_0^2}\right)\!\left(\frac{r^2}{l_0^2}\right)^{\!\!|m|/2}\!L_n^{|m|}\!\left(\frac{r^2}{l_0^2}\right)\\
\label{HO1_Momentum}
K_{nm}^{(HO)}(k)&=(-1)^n2^\frac{1}{2}l_0\left[\frac{n!}{(n+|m|)!}\right]^{1/2}\!\!\exp\!\left(\!-\frac{l_0^2k^2}{2}\right)(l_0k)^{|m|}L_n^{|m|}\!\left(l_0^2k^2\right),
\end{align}
\end{subequations}
$l_0=\left[\hbar/\left(M\omega_0\right)\right]^{1/2}$. Knowledge of these equations paves the way to calculating the miscellaneous properties of the system; for example, root-mean-square radial position
\begin{subequations}\label{RootMeanSquare1}
\begin{align}\label{RootMeanSquare1_Position}
\mathtt{r}_{nm}&=\sqrt{\int_{\mathcal{D}_\rho^{(\mathtt{d})}}r^2\left|\Psi({\bf r})\right|^2d{\bf r}}
\intertext{and wave vector}
\label{RootMeanSquare1_Momentum}
\mathtt{k}_{nm}&=\sqrt{\int_{\mathcal{D}_\gamma^{(\mathtt{d})}}k^2\left|\Phi({\bf k})\right|^2d{\bf k}}
\end{align}
\end{subequations}
values are evaluated as
\begin{subequations}\label{RootMeanSquare2}
\begin{align}\label{RootMeanSquare2_HOB_Position}
\mathtt{r}_{nm}^{(HO{\bf B})}&=l_{eff}\sqrt{2(2n+|m|+1)}\\
\label{RootMeanSquare2_HOB_Momentum}
\mathtt{k}_{nm}^{(HO{\bf B})}&=\frac{1}{l_{eff}}\sqrt{\frac{2n+|m|+1}{2}}\\
\label{RootMeanSquare2_B_Position}
\mathtt{r}_{nm}^{({\bf B})}&=l_B\sqrt{2(2n+|m|+1)}\\
\label{RootMeanSquare2_B_Momentum}
\mathtt{k}_{nm}^{({\bf B})}&=\frac{1}{l_B}\sqrt{\frac{2n+|m|+1}{2}}\\
\label{RootMeanSquare2_HO_Position}
\mathtt{r}_{nm}^{(HO)}&=l_0\sqrt{2n+|m|+1}\\
\label{RootMeanSquare2_HO_Momentum}
\mathtt{k}_{nm}^{(HO)}&=\frac{1}{l_0}\sqrt{2n+|m|+1}
\intertext{with their product being the same for all three configurations:}\label{RootMeanSquare2_Product}
\mathtt{r}_{nm}^{(HO{\bf B})}\mathtt{k}_{nm}^{(HO{\bf B})}&=\mathtt{r}_{nm}^{({\bf B})}\mathtt{k}_{nm}^{({\bf B})}=\mathtt{r}_{nm}^{(HO)}\mathtt{k}_{nm}^{(HO)}=2n+|m|+1.
\end{align}
\end{subequations}

Regarding Shannon entropy, one notes that for the higher lying bands, $n\geq1$, its magnitude in either space can be calculated numerically only but for $n=0$ it is possible to express it analytically:
\begin{subequations}\label{HOB_Shannon1}
\begin{align}\label{HOB_ShannonPosition1}
S_{\rho_{0,m}}^{(HO{\bf B})}&=2\ln l_{eff}+1+\ln\pi+\ln2+\ln(|m|!)+|m|[1-\psi(|m|+1)]\\
\label{HOB_ShannonMomentum1}
S_{\gamma_{0,m}}^{(HO{\bf B})}&=-2\ln l_{eff}+1+\ln\pi-\ln2+\ln(|m|!)+|m|[1-\psi(|m|+1)]
\intertext{with the sum}\label{HOB_ShannonSum1}
S_{t_{0,m}}^{(HO{\bf B})}&=2(1+\ln\pi)+2\ln(|m|!)+2|m|[1-\psi(|m|+1)].
\end{align}
\end{subequations}
Here, $\psi(x)=d[\ln\Gamma(x)]/dx=\Gamma'(x)/\Gamma(x)$ is psi, or digamma function, which is a derivative of the logarithm of the $\Gamma$-function \cite{Abramowitz1}. Last equation shows that the magnetic intensity does not alter the amount of the total information about position and momentum of the particle dwelling in the harmonic potential. Of course, this statement is elementary extended to any orbital \cite{Olendski3}. For the 2D HO, equations~\eref{HOB_ShannonPosition1} and \eref{HOB_ShannonMomentum1} read:
\begin{subequations}\label{HO_Shannon1}
\begin{align}\label{HO_ShannonPosition1}
S_{\rho_{0,m}}^{(HO)}&=2\ln l_0+1+\ln\pi+\ln(|m|!)+|m|[1-\psi(|m|+1)]\\
\label{HO_ShannonMomentum1}
S_{\gamma_{0,m}}^{(HO)}&=-2\ln l_0+1+\ln\pi+\ln(|m|!)+|m|[1-\psi(|m|+1)],
\end{align}
\end{subequations}
whereas their sum does coincide with its fellow from equation~\eref{HOB_ShannonSum1}. The last statement holds true for the unperturbed magnetic field too with the position and momentum components for this configuration being obtained from equations~\eref{HOB_ShannonPosition1} and \eref{HOB_ShannonMomentum1} by replacing $l_{eff}$ by $l_B$:
\begin{subequations}\label{B_Shannon1}
\begin{align}\label{B_ShannonPosition1}
S_{\rho_{0,m}}^{({\bf B})}&=2\ln l_B+1+\ln\pi+\ln2+\ln(|m|!)+|m|[1-\psi(|m|+1)]\\
\label{B_ShannonMomentum1}
S_{\gamma_{0,m}}^{({\bf B})}&=-2\ln l_B+1+\ln\pi-\ln2+\ln(|m|!)+|m|[1-\psi(|m|+1)].
\end{align}
\end{subequations}
Remarkably, for all three geometries, the ground orbital $n=m=0$ does saturate Shannon inequality~\eref{ShannonInequality1}, as it immediately follows from equation~\eref{HOB_ShannonSum1}. It is also important to point out that for the 2D harmonic QD, dimensionless position and momentum functionals of this level are equal and, accordingly, contribute the same amount to the total entropy $S_{t_{00}}$ making it equal to the fundamental limit $2(1+\ln\pi)$:
\begin{subequations}\label{Equality1}
\begin{align}\label{Equality1_1}
S_{\rho_{00}}^{(HO)}-2\ln l_0&=S_{\gamma_{00}}^{(HO)}+2\ln l_0=1+\ln\pi.
\intertext{At the same time, for the uniform magnetic field and its combination with electrostatic HO these measures are different:}
\label{Equality1_2}
S_{\rho_{00}}^{(HO{\bf B})}-2\ln l_{eff}&=S_{\rho_{00}}^{({\bf B})}-2\ln l_B=1+\ln\pi+\ln2\\
\label{Equality1_3}
S_{\gamma_{00}}^{(HO{\bf B})}+2\ln l_{eff}&=S_{\gamma_{00}}^{({\bf B})}+2\ln l_B=1+\ln\pi-\ln2.
\end{align}
\end{subequations}

Similar to their position counterpart of the Volcano ring \cite{Olendski3}, both components of the Fisher information of any orbital of all three geometries depend on the principal quantum index $n$ only:
\begin{subequations}\label{HOB_Fisher1}
\begin{align}\label{HOB_FisherPosition1}
I_{\rho_{nm}}^{(HO{\bf B})}l_{eff}^2&=I_{\rho_{nm}}^{({\bf B})}l_B^2=4n+2\\
\label{HOB_FisherMomentum1}
\frac{I_{\gamma_{nm}}^{(HO{\bf B})}}{l_{eff}^2}&=\frac{I_{\gamma_{nm}}^{({\bf B})}}{l_B^2}=4(4n+2)\\
\label{HO_FisherPositionMomentum1}
I_{\rho_{nm}}^{(HO)}l_0^2&=\frac{I_{\gamma_{nm}}^{(HO)}}{l_0^2}=2(4n+2).
\intertext{As equation~\eqref{HO_FisherPositionMomentum1} manifests, for the field-free HO the dimensionless position and momentum functionals are equal to each other whereas for the configurations with the fields the former ones are four times smaller than the latters. For all three sructures, the products of the two are equal:}
\label{HOB_FisherProduct1}
I_{\rho_{nm}}^{(HO{\bf B})}I_{\gamma_{nm}}^{(HO{\bf B})}&=I_{\rho_{nm}}^{({\bf B})}I_{\gamma_{nm}}^{({\bf B})}=I_{\rho_{nm}}^{(HO)}I_{\gamma_{nm}}^{(HO)}=4(4n+2)^2.
\end{align}
\end{subequations}

Lowest band informational energies are:
\begin{subequations}\label{HOB_Onicescu1}
\begin{align}\label{HOB_OnicescuPosition1}
O_{\rho_{0,m}}^{(HO{\bf B})}&=\frac{1}{2\pi}\frac{1}{l_{eff}^2}\frac{(2|m|)!}{2^{2|m|+1}(|m|!)^2}\\
\label{HOB_OnicescuMomentum1}
O_{\gamma_{0,m}}^{(HO{\bf B})}&=\frac{l_{eff}^2}{2\pi}\frac{(2|m|)!}{2^{2|m|-1}(|m|!)^2}\\
O_{\rho_{0,m}}^{({\bf B})}&=\frac{1}{2\pi}\frac{1}{l_B^2}\frac{(2|m|)!}{2^{2|m|+1}(|m|!)^2}\\
\label{B_OnicescuMomentum1}
O_{\gamma_{0,m}}^{({\bf B})}&=\frac{l_B^2}{2\pi}\frac{(2|m|)!}{2^{2|m|-1}(|m|!)^2}\\
\label{HO_OnicescuPosition1}
O_{\rho_{0,m}}^{(HO)}&=\frac{1}{2\pi}\frac{1}{l_0^2}\frac{(2|m|)!}{2^{2|m|}(|m|!)^2}\\
\label{HO_OnicescuMomentum1}
O_{\gamma_{0,m}}^{(HO)}&=\frac{l_0^2}{2\pi}\frac{(2|m|)!}{2^{2|m|}(|m|!)^2}
\intertext{with their product}\label{HOB_OnicescuProduct1}
O_{\rho_{0,m}}^{(HO{\bf B})}O_{\gamma_{0,m}}^{(HO{\bf B})}&=O_{\rho_{0,m}}^{({\bf B})}O_{\gamma_{0,m}}^{({\bf B})}=O_{\rho_{0,m}}^{(HO)}O_{\gamma_{0,m}}^{(HO)}=\frac{1}{(2\pi)^2}\frac{[(2|m|)!]^2}{2^{4|m|}(|m|!)^4}.
\end{align}
\end{subequations}
For $m=0$, the right-most-side of chain~\eref{HOB_OnicescuProduct1} reaches maximum that is an upper limit from inequality~\eref{OnicescuInequality1}. Similar to the Shannon entropies and Fisher informations, dimensionless HO position and momentum disequilibria  of this orbital are equal to each other:
\begin{subequations}\label{Equality2}
\begin{align}\label{Equality2_1}
O_{\rho_{00}}^{(HO)}l_0^2&=\frac{O_{\gamma_{00}}^{(HO)}}{l_0^2}=\frac{1}{2\pi},
\intertext{and for other configurations, the unitless position functional is four times less its momentum counterpart, as was the case for the Fisher informations too:}
\label{Equality2_2}
O_{\rho_{00}}^{(HO{\bf B})}l_{eff}^2&=O_{\rho_{00}}^{({\bf B})}l_B^2=\frac{1}{4\pi}\\
\label{Equality2_3}
\frac{O_{\gamma_{00}}^{(HO{\bf B})}}{l_{eff}^2}&=\frac{O_{\gamma_{00}}^{({\bf B})}}{l_B^2}=\frac{1}{\pi}.
\end{align}
\end{subequations}
It has to be noted that the informational energies have analytic representations for the higher bands too; e.g., for $n=1$, one has for the position component:
\begin{equation}\label{Onicescu_HigherBand1}
O_{\rho_{1,m}}^{(HO{\bf B})}=\frac{1}{2\pi}\frac{1}{8l_{eff}^2}\frac{(2|m|)!}{[(|m|+1)!]^2}\left(\frac{|m|^2}{34^{|m|}}+\frac{|m|}{54^{|m|}}+\frac{1}{82^{2|m|+2}}\right),
\end{equation}
but with the growth of $n$, they become increasingly unwieldy and hard to comprehend.

\subsection{Shannon entropy}\label{Sec_Shannon}
\begin{figure}
\centering
\includegraphics[width=\columnwidth]{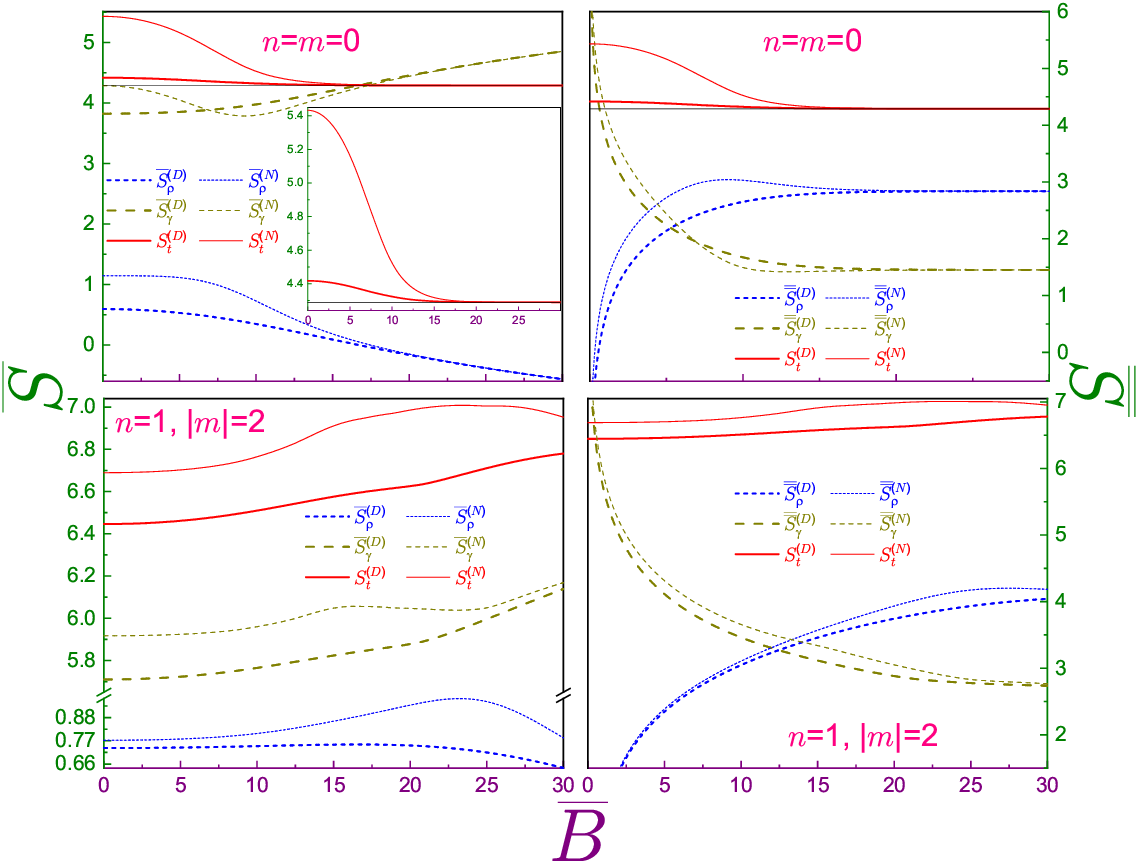}
\caption{\label{Fig_Shannon}
Normalized Shannon entropies $\overline{S}$ (left plots) and $\overline{\overline{S}}$ (right figures) of the Dirichlet (superscript $D$, thick lines) and Neumann ($N$, thin curves) QDs as functions of the dimensionless magnetic field $\overline{B}$ where upper (lower) panels are for $n=m=0$ ($n=1$, $|m|=2$) orbital. Dot dependencies depict position components $S_\rho$, dashed lines are for the momentum functionals $S_\gamma$, and solid ones exhibit $S_t$. Horizontal dark lines in the upper subplots are a fundamental limit $2(1+\ln\pi)=4.2894\ldots$ from the right-hand side of inequality~\eqref{ShannonInequality1}. Inset in the upper left window provides an enlarged view of the Dirichlet  and Neumann entropies $S_t$. Note different vertical ranges and scales of each subplot and a line break from $0.97$ to $5.65$ in the lower left graph.} 
\end{figure}

Figure~\ref{Fig_Shannon} shows Shannon entropy evolution with the dimensionless intensity $\overline{B}$ for the $n=m=0$ and $n=1$, $|m|=2$ states. Before, it was shown \cite{Olendski1} that, at the field-free geometry, this measure in either space has higher magnitude for each Neumann level what physically implies more knowledge about the corresponding property of the Dirichlet dot. Obviously, the sum of the two entropies $S_t$ inherits this property too; for example, its $n=m=0$ values are $4.4174$ and $5.4327$ for the Dirichlet and Neumann BC, respectively \cite{Olendski1}. Applied field squeezes the position waveform and spreads the momentum one, as shown in figures~\ref{Fig_PositionFunctions} and \ref{Fig_MomentumFunctions}, respectively, what means that it leads to the gain or loss of the corresponding information. Accordingly, position (momentum) component $\overline{S}$ that is measured in electric units decreases (increases) with $\overline{B}$ in such a way that the total $n=m=0$ entropy monotonically declines and at the high intensities, due to the negligible influence of the interface, approaches, regardless of the surface requirement, a fundamental limit $2(1+\ln\pi)=4.2894\ldots$. This change of the sum $S_\rho+S_\gamma$ is a peculiar feature of any BC hard-wall potential that tells it from the harmonic confinement for which, as discussed above, equation~\eref{HOB_ShannonSum1}, $S_t$ does not depend on the field. Upper panels show that the Dirichlet sum nears at the smaller $\overline{B}$ the restraint from the right-hand side of inequality~\eref{ShannonInequality1} what is naturally explained by the opposite actions of the Neumann interface and the magnetic intensity. Since at $B=0$ the total Neumann entropy  is further from the fundamental limit, the absolute value of its speed of change with the field is greater as compared to the Dirichlet dot: 
\begin{equation}\label{ShannonInequality2}
\left|\frac{\partial S_{t_{00}}^{(N)}}{\partial\overline{B}}\right|>\left|\frac{\partial S_{t_{00}}^{(D)}}{\partial\overline{B}}\right|,
\end{equation}
what is vividly underlined in the inset of the upper left panel. Measured in the magnetic units, the entropy almost ceases to vary at the strong $\overline{B}$, as upper right window demonstrates: at $\overline{B}\gtrsim20$, both Dirichlet and Neumann position functionals $\overline{\overline{S}}_{\rho_{00}}$ practically coincide with each other and both come closer and closer to the unperturbed-field value $1+\ln\pi+\ln2=2.837\ldots$ from equation~\eref{Equality1_2} with their momentum fellows $\overline{\overline{S}}_{\gamma_{00}}$ tending almost simultaneously to $1+\ln\pi-\ln2=1.451\ldots$, equation~\eref{Equality1_3}. In this regime, the particle is so strongly squeezed to the center by the applied field that it does not 'see' the dot surface and, accordingly, is not influenced by it; hence, the distinction between the two BCs vanishes. Let us also note that the Neumann momentum component $\overline{S}_{\gamma_{00}}$ is a nonmonotonic function of the field; e.g., its decrease at the small and moderate intensities leads to the intersection at $\overline{B}\approx6.9$ with its Dirichlet counterpart and at any greater field the former is smaller than the latter. Since simultaneously, $\overline{S}_{\rho_{00}}^{(D)}<\overline{S}_{\rho_{00}}^{(N)}$, the total entropy at any BC approaches its limit faster than its position or momentum component.

Nonmonotonicity of the entropies becomes more conspicuous for the levels with larger $n$ and $|m|$ when the zero-field position wave function is distributed closer to the circumference and/or has more nodes. This is exemplified by the lower windows which show the Shannon functionals of $n=1$, $|m|=2$ states. For this orbital, both position components $\overline{S}_\rho$ initially grow with the field and only at the strong enough intensities reverse the behaviour. Neumann total entropy $S_{t_{1,2}}^{(N)}$ is a nonmonotonic function of the field too staying, however, at any  $\overline{B}$ above its Dirichlet partner what again is explained by the fact that the Dirichlet wall helps the field to push the electron to the center whereas the Neumann surface acts in the opposite direction diminishing in this way the total amount of information about the particle. Since for these levels the influence of the QD surface is more pronounced, higher intensities $\overline{B}$ are needed to reach the unperturbed-field limits; in particular, this asymptotic approach is not shown in the range of figure~\ref{Fig_Shannon}.

\subsection{Fisher information}\label{Sec_Fisher}
\begin{figure}
\centering
\includegraphics[width=\columnwidth]{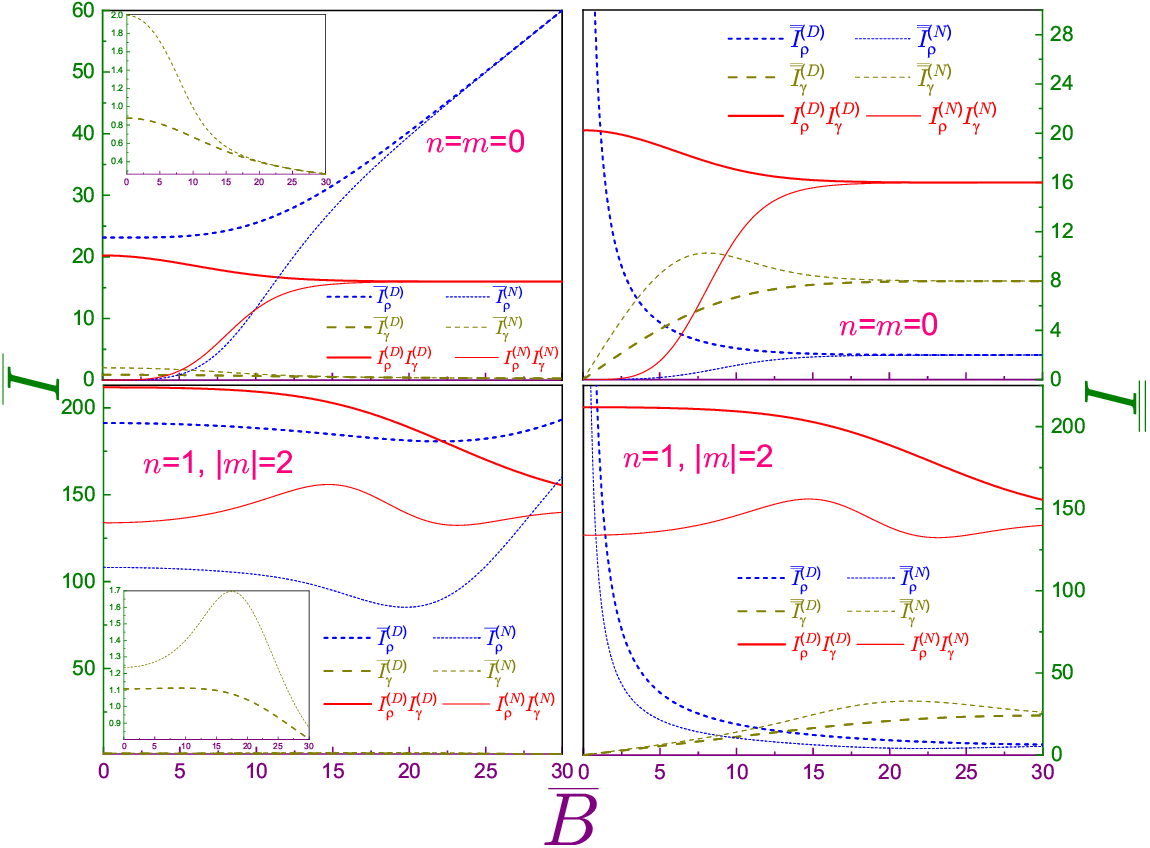}
\caption{\label{Fig_Fisher}
Normalized Fisher informations $\overline{I}$ (left plots) and $\overline{\overline{I}}$ (right figures) of the Dirichlet (superscript $D$, thick lines) and Neumann ($N$, thin curves) QDs as functions of the dimensionless magnetic field $\overline{B}$ where upper (lower) panels are for $n=m=0$ ($n=1$, $|m|=2$) orbital. Dot dependencies depict position components $I_\rho$, dashed lines are for the momentum functionals $I_\gamma$, and solid ones exhibit the product $I_\rho I_\gamma$. Insets in the left windows provide enlarged views of the Dirichlet  and Neumann momentum informations $\overline{I}_\gamma$. Note different vertical ranges and scales of each subplot.} 
\end{figure}

Field-free Neumann position wave function of the $n=m=0$ state is just a constant
\begin{equation}\label{ZeroFieldNeumannPositionFunction1}
\left.\Psi_{00}^{(N)}({\bf r})\right|_{B=0}=\frac{1}{\pi^{1/2}a},
\end{equation}
what automatically zeroes the associated Fisher information with its dimensionless momentum counterpart being equal to two \cite{Olendski1}. Normalized Dirichlet position component of any circularly symmetric level in the same regime is just a quadruple of the square of the corresponding zero of the Bessel function $j_{0,n+1}$ \cite{Olendski1}. By squeezing the particle to the center, applied field simultaneously increases the slope of the radial position waveform resulting in the increase of the related Fisher components $\overline{I}_{\rho_{00}}$. This monotonic growth is depicted in the left upper panel of figure~\ref{Fig_Fisher} with the Dirichlet curve at any $\overline{B}$ lying above the Neumann dependence. Upon the magnetic influence, the momentum function  flattens with its oscillations being more and more subdued what is a primary reason of the monotonic decrease of the momentum functionals $\overline{I}_{\gamma_{00}}$ with the Dirichlet component, contrary to the position measures, being always smaller, as the inset of the window emphasizes. Interaction of the position and momentum swingings results in the overall oscillating power of the orbital $|nm\rangle$ that is quantitatively expressed by the product $I_\rho I_\gamma$. As the corresponding solid lines manifest, Dirichlet level exhibits more swaying abilities with the difference between the two BCs, however, decreasing with the field what, as before, is explained by the diminishing influence of the wall. Switching to the magnetic units, one observes that at the strong $\overline{B}$ the position $\overline{\overline{I}}_{\rho_{00}}$ and momentum $\overline{\overline{I}}_{\gamma_{00}}$ parts tend to the unperturbed-field values, which, according to equations~\eref{HOB_FisherPosition1} and \eref{HOB_FisherMomentum1}, are $2$ and $8$, respectively.

The orbitals with nonzero indices exhibit nonmonotonic $\overline{I}_\rho$ and $\overline{I}_\gamma$ dependencies on the field, as the lower left panel shows: for either BC, the position (momentum) component has a minimum (maximum) on the $\overline{B}$-axis. Their interplay yields two Neumann extrema whereas the product of the two Dirichlet parts $I_{\rho_{1,2}}^{(D)}I_{\gamma_{1,2}}^{(D)}$ is a monotonic function that at the large fields tends from above to its unperturbed-field value of $144$, equation~\eref{HOB_FisherProduct1}. Since at these quantum numbers the wall influence is stronger than at $n=m=0$, this asymptotic approach takes place at the higher fields and only its initial stage is shown in the lower subplots. Neumann overall oscillating power remains weaker than its Dirichlet fellow, in particular, it reaches the above-mentioned limit from below. On the magnetic scale, two position components $\overline{\overline{I}}_\rho$ at the growing $\overline{B}$ come closer and closer to each other simultaneously closing on the value of six, as it follows from equation~\eref{HOB_FisherPosition1}. The same is true for the momentum parts with their asymptote of $24$, equation~\eref{HOB_FisherMomentum1}.

\subsection{Disequilibrium}\label{Sec_Onicescu}
\begin{figure}
\centering
\includegraphics[width=\columnwidth]{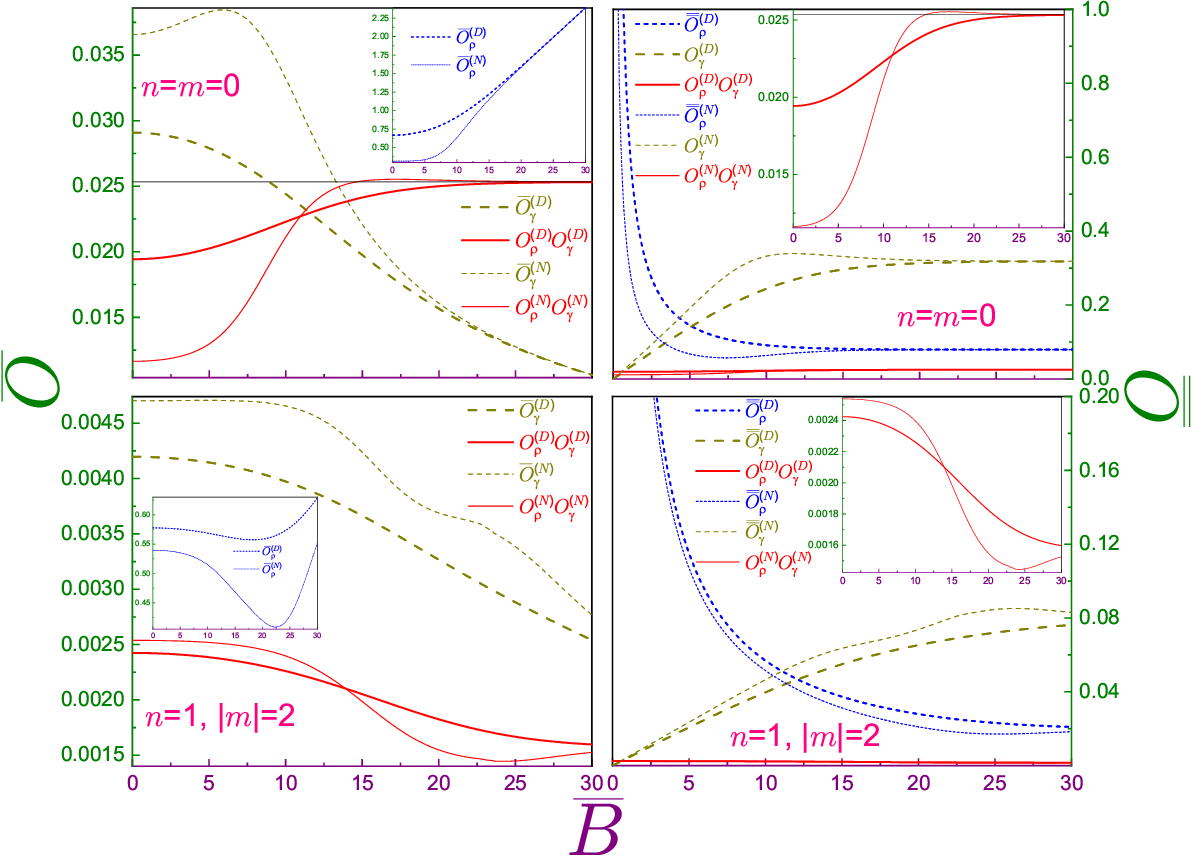}
\caption{\label{Fig_Onicescu}
The same as in figure~\ref{Fig_Fisher} but for the informational energies. Insets in the left windows  exhibit the Dirichlet  and Neumann position functionals $\overline{O}_\rho$ and in the right panels they provide enlarged views of the products $O_\rho O_\gamma$. Horizontal dark lines in the upper left subplot and in the inset of the upper right panel are the value $(2\pi)^{-2}=0.02533\ldots$ from the right-hand side of inequality~\eqref{OnicescuInequality1}. Note different vertical ranges and scales of each subplot.} 
\end{figure}

Zero-field position Neumann waveform from equation~\eref{ZeroFieldNeumannPositionFunction1} not only provides the minimum of information among all the corresponding levels of both BC types (what, in this regime, maximizes its entropy $\overline{S}_\rho$) and not only is the least oscillating dependence (what results in the minimal, i.e., zero, Fisher information $I_\rho$ among all Dirichlet and Neumann states) but also, due to its uniformity, possesses the lowest position informational energy, $\left.\overline{O}_{\rho_{00}}^{(N)}\right|_{B=0}=1/\pi=0.3183\ldots$ \cite{Olendski1}. Applied field destroys homogeneity and opens up in this way three processes: first, it provides more information about particle location what causes the decrease of the Shannon entropy $\overline{S}_\rho$, as discussed in Sec.~\ref{Sec_Shannon}; secondly, position function acquires some oscillating structure thus leading to the growth of the Fisher information $\overline{I}_\rho$ what was addressed in the previous subsection; and, relevant to this part of our discussion, it deviates the function from the uniformity leading to the growth of the disequilibrium $\overline{O}_{\rho_{00}}^{(N)}$. Inset in the left upper panel of figure~\ref{Fig_Onicescu} shows that this increase is a monotonic one and at strong enough intensities $\overline{B}$ it comes closer to its Dirichlet counterpart (which also is a monotonic function) staying, however, always below $\overline{O}_{\rho_{00}}^{(D)}$ and both of them tend at $B\rightarrow\infty$ to the uniform field value from equation~\eref{Equality2_2}, i.e., become linear functions of $\overline{B}$. Momentum component $\overline{O}_{\gamma_{00}}^{(D)}$ steadily decreases with the field and its Neumann fellow passes through the maximum after which it comes closer and closer to the Dirichlet disequilibrium with both of them obeying in this regime the inverse field dependence. Overall deviation from the uniformity is quantified by the product of the position and momentum functionals $O_\rho O_\gamma$ and for the Dirichlet $|00\rangle$ orbital it monotonically approaches from below the limit $1/(2\pi)^2=0.02533\ldots$, in accordance with inequality~\eref{OnicescuInequality1}. However, the most remarkable is the behaviour of the total Neumann disequilibrium; namely, at its growth with the field, it overpasses at $\overline{B}\approx10.95$ the Dirichlet product and, even more importantly, at $\overline{B}\gtrsim14.6$ it turns greater than the right-hand side of inequality~\eref{OnicescuInequality1}. Thus, our analysis of the Neumann QD in the field refutes the universality of this relation. At further growth of the  magnetic intensity, the product $O_{\rho_{00}}^{(N)}O_{\gamma_{00}}^{(N)}$ reaches at $\overline{B}\approx17.2$ a wide maximum of $0.025528\ldots$ and then smoothly decreases to $(2\pi)^{-2}$. In terms of the magnetic units, the functionals at quite strong fields almost stop changing: they stay in a very close vicinity of the values from equations~\eref{Equality2_2} and \eref{Equality2_3} since, as was already stated several times above, the wall influence becomes negligible.

Lower left panel of figure~\ref{Fig_Onicescu} reveals that the functional $\overline{O}_\rho$ for either QD exhibits a nonmonotonic dependence on the dimensionless field with the BC dependent minimum. It also shows that the overall Dirichlet disequilibrium is a steady decreasing function with its Neumann counterpart having minimum on $\overline{B}$ axis. Similar to the previous measures, all informational energies of this level reach the uniform field limit at the higher $\overline{B}$ that lies outside the range of figure~\ref{Fig_Onicescu}.

\section{Concluding remarks}\label{Sec_Conclusions}
A race for theoretical scrutinization of quantum-information measures does not show any signs of slowing down. These efforts get more impetus due to, among others, experimental advances in detecting, probing and evaluating them \cite{Lukin1,Niknam1,Islam1,Kaufman1,Brydges1}. Present research contributes to that part of this huge investigation project that is devoted to the analysis of the magnetic field influence on them and the role played by the different BCs in this effect. Considered by us measures describe different facets of position and momentum probability distributions: Shannon entropy quantifies the information that is missing, Fisher information expresses numerically the oscillating power of the corresponding orbital and the value of the informational energy illustrates the deviation from the uniformity. We have shown that, by pushing the electron away from itself, the Dirichlet interface helps the increasing field to squeeze the particle to the dot center whereas the Neumann surface attracts the corpuscle and, accordingly, impedes the magnetic intensity efforts. This results in different spectra for the different BCs; in particular, characteristic feature of the Neumann magnetic energies is their crossings. We have shown that with the increasing intensity $\overline{B}$, the Shannon measures of the Dirichlet QD approach faster their counterparts of the unperturbed magnetic field what is just explained by the mentioned above different roles played by the two edge requirements. Among other results, we would like to point out that the disequilibrium inequality~\eref{OnicescuInequality1} needs additional rethinking since it is violated by the Neumann QD in the field.

Dirichlet, equation~\eref{BC1_Dirichlet}, and Neumann, equation~\eref{BC1_Neumann}, surface requirements are particular limiting cases $\Lambda=0$ and $\Lambda=\infty$, respectively, of the linear relation (Robin BC \cite{Gustafson1}) between the position waveform and its spatial derivative at the surface $\cal S$:
\begin{equation}\label{Robin1}
\left.{\bf n}{\bm\nabla}_{\bf r}\Psi(\bf r)\right|_{\cal S}=\frac{1}{\Lambda}\Psi(\bf r)|_{\cal S},
\end{equation}
where the extrapolation length $\Lambda$ in general is a function of the location on the interface. Position dependencies $\Psi_{nm}^{(R)}({\bf r})$ and energy spectra $E_{nm}^{(R)}$ of the Robin disc with positive, negative and even complex $\Lambda$ in the field $B$ are known \cite{Olendski8} but apparently nobody considered momentum parts and associated quantum-information measures in both spaces. Their analysis might help to clarify the peculiarities of transformation from the Dirichlet to Neumann BC, in particular, to shed the light on the properties of the disequilibrium inequality.

\end{document}